\documentclass[oldversion]{aa}

\usepackage{graphicx}
\usepackage{graphpap}
\usepackage{epsf}
\usepackage{amssymb}
\usepackage{color}

\usepackage{natbib}

\def\FIG #1 #2 [#3] #4\par{%
  \begin{figure}\begin{center}%
    \includegraphics*[#3]{#2}%
    \\
    \caption{#4}%
    \label{#1}%
  \end{center}\end{figure}%
}
\def\FIGG #1 #2 #3 [#4] #5\par{%
  \begin{figure*}[ht]\begin{center}%
               \includegraphics*[#4]{#2}
               \includegraphics*[#4]{#3}%}%
    \caption{#5}%
    \label{#1}%
   \end{center}\end{figure*}%
}
\def\FIGth #1 #2 #3 #4 [#5] #6\par{%
  \begin{figure*}[ht]\begin{center}%
        \includegraphics[#5]{#2} \\
        \includegraphics[#5]{#3}
        \includegraphics[#5]{#4}
        \caption{#6}
        \label{#1}
   \end{center}\end{figure*}
}
\def\FIGfo #1 #2 #3 #4 #5 [#6] #7\par{%
  \begin{figure*}[ht]\begin{center}%
        \includegraphics[#6]{#2}
        \includegraphics[#6]{#3} \\
        \includegraphics[#6]{#4}
        \includegraphics[#6]{#5}
        \caption{#7}
        \label{#1}
   \end{center}\end{figure*}
}
\def\FIGsi #1 #2 #3 #4 #5 #6 #7 [#8] #9\par{%
  \begin{figure*}[ht]\begin{center}%
        \includegraphics*[#8]{#2}
        \includegraphics*[#8]{#3}\\
        \includegraphics*[#8]{#4}
        \includegraphics*[#8]{#5}\\
        \includegraphics*[#8]{#6}
        \includegraphics*[#8]{#7}
        \caption{#9}
        \label{#1}
    \end{center}\end{figure*}
}

\def\rfig#1{Fig. \ref{#1}}

\newcommand{\mt}[1]{\mathrm{#1}}
\newcommand{\sect}[1]{Section~\ref{#1}}
 
\def\lvm{\leavevmode\hbox to\parindent{\hfill}}
\def\req#1{(\ref{#1})}

\def\d{\partial}   
\def\dT{\partial T}
\def\dt{\partial t}
\def\du{\partial u}
\def\dr{\partial r}
\def\dm{\partial m}

\def\dE{\partial E}

\def\dX{\partial X}

\def\eps{\varepsilon}

\def\BE{\begin{equation}}
\def\EE{\end{equation}}
\def\BA{\begin{array}}
\def\EA{\end{array}}
\def\BAN{\begin{eqnarray*}}
\def\EAN{\end{eqnarray*}}

\def\fun#1#2{\lower3.6pt\vbox{\baselineskip0pt\lineskip.9pt
\ialign{$\mathsurround=0pt#1\hfil##\hfil$\crcr#2\crcr\sim\crcr}}}

\def\ltsima{$\; \buildrel < \over \sim \;$}
\def\ltsim{\lower.5ex\hbox{\ltsima}}
\def\gtsima{$\; \buildrel > \over \sim \;$}
\def\gtsim{\lower.5ex\hbox{\gtsima}}

\def\Ecr{E_\mt{CR}}
\def\Pcr{P_\mt{CR}}

\def\qcr{q_\mt{CR}}
\def\kcr{\kappa_\mt{CR}}
\def\ecr{\epsilon_\mt{esc}}

\def\kcrd{\;\mt{cm}^2\mt{s}^{-1}}

\usepackage{txfonts}
\bibpunct{(}{)}{,}{a}{}{,}

\begin{document}

\title{Modeling supernova remnants: effects of diffusive cosmic-ray acceleration on the evolution, application to observations}
\titlerunning{Modeling CR acceleration in SNRs}

\author{D.~Kosenko\inst{1} \and S. I. Blinnikov\inst{2,3,4} \and J.~Vink\inst{1}}
   \institute{
Astronomical Institute Utrecht, Utrecht University, P.O. Box 80000, 3508TA Utrecht, The Netherlands \\
    \email{D.Kosenko@uu.nl}
         \and
ITEP, Moscow,  117218, Russia
	\and
SAI, Moscow, 119992, Russia
	\and
IPMU, Tokyo Universiy, Kashiwa, 277-8583, Japan
             }

\authorrunning{Kosenko et al.}
\date{Received ...; accepted ... }

\abstract{
We present numerical models for supernova remnant evolution, using a new version of the hydrodynamical code SUPREMNA. We added cosmic ray diffusion equation to the code scheme, employing two-fluid approximation. We investigate the dynamics of the simulated supernova remnants with different values of cosmic ray acceleration efficiency and diffusion coefficient. We compare the numerical models with observational data of Tycho's and SN1006 supernova remnants. We find models which reproduce the observed locations of the blast wave, contact discontinuity, and reverse shock for the both remnants, thus allowing us to estimate the contribution of cosmic ray particles into total pressure and cosmic-ray energy losses in these supernova remnants. We derive that the energy losses due to cosmic rays escape in Tycho's supernova remnant are 10-20\% of the kinetic energy flux and 20-50\% in SN1006.
}

%\keywords{cosmic ray acceleration, numerical methods, galactic supernova remnants}

\maketitle

\section{Introduction}

Clear evidence for effective acceleration of cosmic-ray (CR) particles in young supernova remnants (SNR) stems from TeV observations of the Galactic sources by HESS \citep{aharonian05,aharonian06}, CANGAROO-II \citep{katagiri05}, MAGIC \citep{magic07}, SHALON \citep{sinitsyna09}, VERITAS \citep{veritas09, veritas10}.
In addition, discoveries of non-thermal X-ray emission from SNRs  \citep[see a review by][]{reynolds08} point to the presence of electrons accelerated to TeV energies in supernova remnants (SNR).

In the last decades a number of numerical methods which account for CR acceleration in simulations of supernova remnants were developed. An extensive review of some of these techniques can be found in \citet{malkovdrury01} and \citet{caprioli10}.
Nearby Galactic SNRs provide an excellent opportunity to test these models and to study the efficiency of CR acceleration processes. For example,  the proximity of the blast wave (BW) to contact discontinuity (CD) in Tycho SNR measured by  \citet{warren05} is inconsistent with adiabatic hydrodynamic models of SNR evolution, and can be explained only if cosmic-ray acceleration of the particles occurs at the forward shock.  The similar evidence was presented for SN1006 supernova remnant by \citet{chenai08} and \citet{miceli09}.
%recent measurements of the blast wave (BW), contact discontinuity (CD) and reverse shock (RS) locations in Tycho \citep{warren05} and SN1006 \citep[][only BW and CD locations]{chenai08, miceli09} and numerical simulations of these remnants confirm that it is not possible to explain the observation without involvement of the CR acceleration and additional energy losses in these remnants.
%\rr{It was not obvious in reading the second paragraph of the Introduction how the authors, and presumably others, arrived at this conclusion, and how robust the conclusion was. Some discussion along these lines is needed. In particular, it appears that these conclusions are based on hydrodynamical simulations, which must immediately raise a question about the importance of the magnetic field in providing realistic models against which to compare observations.}

Both these objects were already used in testing of the numerical and analytical models \citep[e.g.][]{ellison01, ellison07, volk08} of CR acceleration in SNRs.  2D simulations of Tycho's SNR evolution and investigation of Rayleigh-Taylor instability development with the gas adiabatic index values down to $\gamma = 1.1$ were performed by \citet{wang10}.  3D hydrodynamical modeling of Tycho was conducted by \citet{ferrand10}.

The effects of the shock modification by cosmic rays in Kepler SNR were studied by \citet{decourchelle00}, who modeled the X-ray spectra using a non-linear non-equilibrium ionization method. A more detailed study of the acceleration effects on the thermal emission from the shocked supernova ejecta was conducted by \citet{patnaude10}.

In this study we present hydrodynamical (HD) simulations of supernova remnant evolution with the account for diffusive cosmic-ray acceleration.  We introduced a CR diffusion equation into the numerical code {\sc supremna}, developed by \citet{sorokina04, kosenko06}. This code calculates the evolution of a supernova remnant assuming spherical symmetry and taking into account time-dependent ionization and thermal conduction. To include the effects of CR acceleration into the scheme, we apply a two-fluid approximation, i.e. we introduce a CR diffusion equation into the system of hydrodynamical equations.

Employing this renewed package we have created sets of hydrodynamical models with different values of CR-related parameters.  We compare the results of our simulations with the observations of Tycho's and SN1006 supernova remnants.

The paper is structured as follows. In \sect{eqs} we describe the basic equations we use for the simulations, in \sect{runs} we show the results of modeling. We compare our models with the observations in the \sect{obs}. We summarize the results in \sect{results}, discuss them in \sect{discussion} and conclude by \sect{conclusion}.

\section{Basic equations and method}
\label{eqs}
\subsection{Code description}
For modeling the evolution of supernova remnants, we employ the hydrodynamical code {\sc supremna}, which was introduced  by \citet{sorokina04}. The method accounts for electron thermal conduction and includes self-consistent calculations of time-dependent ionization processes. The electron and ion temperature equilibration processes are parametrized.

The code uses an implicit Lagrangean formulation for one dimensional spherical-symmetrical geometry. The hydrodynamical evolution of the remnant is coupled with the system of kinetic equations of ionization balance to calculate self-consistently non-equillibrium ionization state of the shocked plasma. Ion and electron temperatures are treated separately taking into account electron thermal conduction (see Appendix~\ref{equations}).

\if !=
To solve the system of equations~\req{velo} -- \req{dotX}, they developed an implicit finite-difference method,
which is based on the method of lines for the equations of hydrodynamics, when the spatial derivatives are approximated by finite differences, and the emerging system of ordinary differential equations is solved by Gear-type multistep predictor-corrector methods
%(Gear 1971; see also Arushanyan, Zaletkin 1990).
% proceeds
The heat exchange processed in the approximation of thermal conduction according to
Eq.\req{tpe}, which is also solved by the method of lines consistent with hydrodynamics.
%Each of these quantities is explained in details in  \citet{sorokina04}.
\fi

To describe the effects of collisionless energy exchange, \citet{sorokina04}  introduced a parameter $q_i$ ($0<q_i<1$) which specifies a fraction of artificial viscosity $Q$, added to the pressure of ions in the equations \req{accel},~\req{tpi} and plays a role of a source term (the details are presented in the Appendix). If only the collisional exchange is taken into account, then $q_i=(1.0-m_e/m_p)$ and the standard system of equations with the heating of just ions at the front is being solved.

%The von Neumann artificial viscosity term is defined as follows
The artificial viscosity \citep{richmyer67} term is defined as follows
\BE
Q = \left\{
\begin{array}{cc}
  A_q\rho(\Delta u)^2 & \mt{if} \; (\Delta u)<0\label{artvisc} \\
  0                                  & \mt{otherwise},  \\
\end{array} \right.
\label{artvisq}
\EE
with dimensionless parameter $A_q = 2$ and $\Delta u$ --- velocity difference at neighboring mesh points.
%\dk{end of copy-paste}

\subsection{Cosmic-ray diffusion equation}
In order to tailor the scheme for simulations of SNRs we need to take into account cosmic-ray (CR) diffusion. We used two-fluid approximation  \citep[e.g.][]{kang90, ko95, malkovdrury01, blasi02, blasi04, wagner2006, zirakashvili10} and introduced additional CR diffusion equation into the (\ref{velo})-(\ref{tpi}) set.

The one-dimensional CR diffusion equation in the Eulerian frame for the plane-parallel case %\citep[e.g. equation (2.3a) from ][]{kang90}
reads
\BE
{\d \Ecr\over\dt}+{\d(u\Ecr)\over\dr} - {\d\over\dr}\left(\kcr {\d \Ecr\over\dr}\right) + \Pcr{\du\over\dr} = \Theta,
\label{crtrans}
\EE
where $\Pcr = (\gamma_\mt{CR} - 1)\Ecr$ is CR pressure, $\Ecr$ --- CR energy density, $\Theta$ is an external source of CR energy injection, $\kcr$ --- diffusion coefficient. We assume equation of state for CR matter with the fixed adiabatic exponent of $\gamma_\mt{CR} = 4/3$,

We generate relativistic particles pressure $\Pcr$ from the artificial viscosity $Q$ term by introducing a parameter $\qcr$, that regulates the injection of CR particles. Thus we define the source term as $\Theta = \qcr Q (\du/\dr)$ (see Appendix~\ref{artviscapp}).

Note that a somewhat similar method was used by \citet{zank93}, where the introduction of a source term into CR diffusion equation was performed via a ``thermal leakage'' mechanism. Particle distribution function was divided in two parts, those particles that have momentum higher than a certain value $p_0$ were treated as CR which propagate according to the CR diffusion equation. Thermal particles are energized due to the adiabatic compression or anomalous heating within a subshock. 

In our study we are not concerned with microphysics of the generation and escape of the energetic particles, but more with hydrodynamical consequences of the acceleration.

After transformation of the equation~(\ref{crtrans}) to our adopted Lagrangian frame, we get
\begin{eqnarray}
% {D\Ecr\over\dt}  = & -  &(\Ecr + \Pcr)4\pi\rho{\d(r^2u)\over\dm} -u\,4\pi r^2\rho{\d\Pcr\over\dm}\nonumber\\ % LOL
{D\Ecr\over Dt}  = & -  &(\Ecr + \Pcr)\,4\pi\rho{\d(r^2u)\over\dm}\nonumber\\
% -u\,4\pi r^2\rho{\d\Ecr\over\dm}
 & +  &4\pi\rho{\d\over\dm}\left(r^2\,F_\mt{CR}\right) - 4\pi\rho{\d(r^2u)\over\dm} \qcr Q,
\label{lagcrtrans}
\end{eqnarray}
where $D\Ecr/Dt = \d\Ecr/\dt + 4\pi r^2u\rho(\d\Ecr/\dm)$.

We treat cosmic-ray flux $F_\mt{CR}$ in the same manner as it is done for thermal electron conduction in \citet{sorokina04}.
\BE
F_\mt{CR}  =  {- \kcr \nabla \Ecr \over{1 + \left| \kcr \nabla \Ecr \right|/F_\mt{sat\; CR}}},
\label{crflux}
\EE
where $F_\mt{sat\; CR} = c\Pcr/2$ ($c$ --- speed of light) is saturated value of the cosmic-ray flux (the Eddington approximation). At this stage we assume a constant diffusion coefficient $\kcr$, as we do not have information on the spectrum of cosmic rays. 

In reality the diffusion will depend on particle energy. In that case the diffusion coefficient used by us should be regarded as an pressure weighted mean value. Moreover, according to non-linear cosmic ray acceleration theory, efficient acceleration leads to hard spectra, in which case most of the energy, and hence pressure, is contained by the particles with the highest energies. If the maximum energy is around $10^{15}$ eV we expect a typical diffusion coefficient, under the assumption of Bohm-diffusion, of $\kcr=1/3 r_g c=3\times10^{26} (B/100\, \mu \mt{G})^{-1} (E/10^{15}\,\mt{eV})\; \mt{cm}^2\mt{s}^{-1}$ ($r_g$ --- gyroradius, $B$ --- magnetic field, $E$ --- energy of the particle). The largest role of the diffusion coefficient for our calculations is in the role of CR escape, as it drains energy from the plasma.

And finally, we alter the equation (\ref{accel}) by adding a component of relativistic particle pressure $\Pcr$ in such a way that
\BE
 {\du\over\dt} = -4\pi r^2\,{\d (P_e+P_i+\Pcr)\over\dm}-{Gm\over r^2}.
  \label{craccel}
\EE

The entire system of equations (\ref{velo}), (\ref{craccel}), (\ref{tpe}), (\ref{tpi}) and (\ref{lagcrtrans}) describes the system, where electrons, ions and cosmic-ray components are treated independently. The contributions to the pressure of these three components are governed by two free parameters $q_\mt{i}$ and $\qcr$.

\section{Numerical models}
\label{runs}
We created a library of numerical models with various sets of parameters $\qcr$ and $\kcr$.  In the simulations we used a delayed-detonation thermonuclear explosion model with $E = 1.4\times10^{51}$ erg \citep{woosley07}, initially the CR pressure is $P^0_\mt{CR} = 10^{-10}P_i$, and the temperature of the homogeneous ambient medium is $T_0 = 10^4$ K.  Using the artificial viscosity source term, we turn on CR generation only at the forward shock.

We considered two sets of models of supernova remnants. Parameters for one set (Tycho's case) were taken similar to Tycho's SNR: the remnant is surrounded with homogeneous circumstellar matter of density $\rho_0 =10^{-24}\;\mt{g\,cm}^{-3}$ and the age of the system is 440 years. Note, that there are indications, that the real ambient density in Tycho vicinity is lower. For example,  \citet{katsuda10} from proper motion measurements found that $n_0\lesssim 0.2\;\mt{cm}^{-3}$, thus $\rho_0 \lesssim 0.4\times 10^{-24}\;\mt{g\,cm}^{-3}$. Nevertheless, taking into account that we consider over-energetic initial explosion and that radius of the remnant $R \propto (E/\rho_0)^{1/5}$ is a weak function of ambient density, we assume that our input parameters match Tycho's SNR.

Examples  of hydrodynamical profiles of a few of these models are presented in \rfig{c440}. Top row shows the simulation with $\qcr = 0.0$, middle row --- $\qcr = 0.7$, $\kcr = 10^{25}\,\mt{cm}^2\mt{s}^{-1}$;  bottom row --- $\qcr = 0.99$, $\kcr = 10^{26}\,\mt{cm}^2\mt{s}^{-1}$. Left panels: density (black solid, scale at the left-hand side), ion pressure (blue dash-dotted, scale at the right-hand side), CR pressure (blue dashed, scale at the right-hand side). Right panels: solid line shows velocity profile (scale at the left-hand side), dashed line --- electron temperature profiles, dashed-dotted line --- ion temperature profiles (scale at the right-hand side).

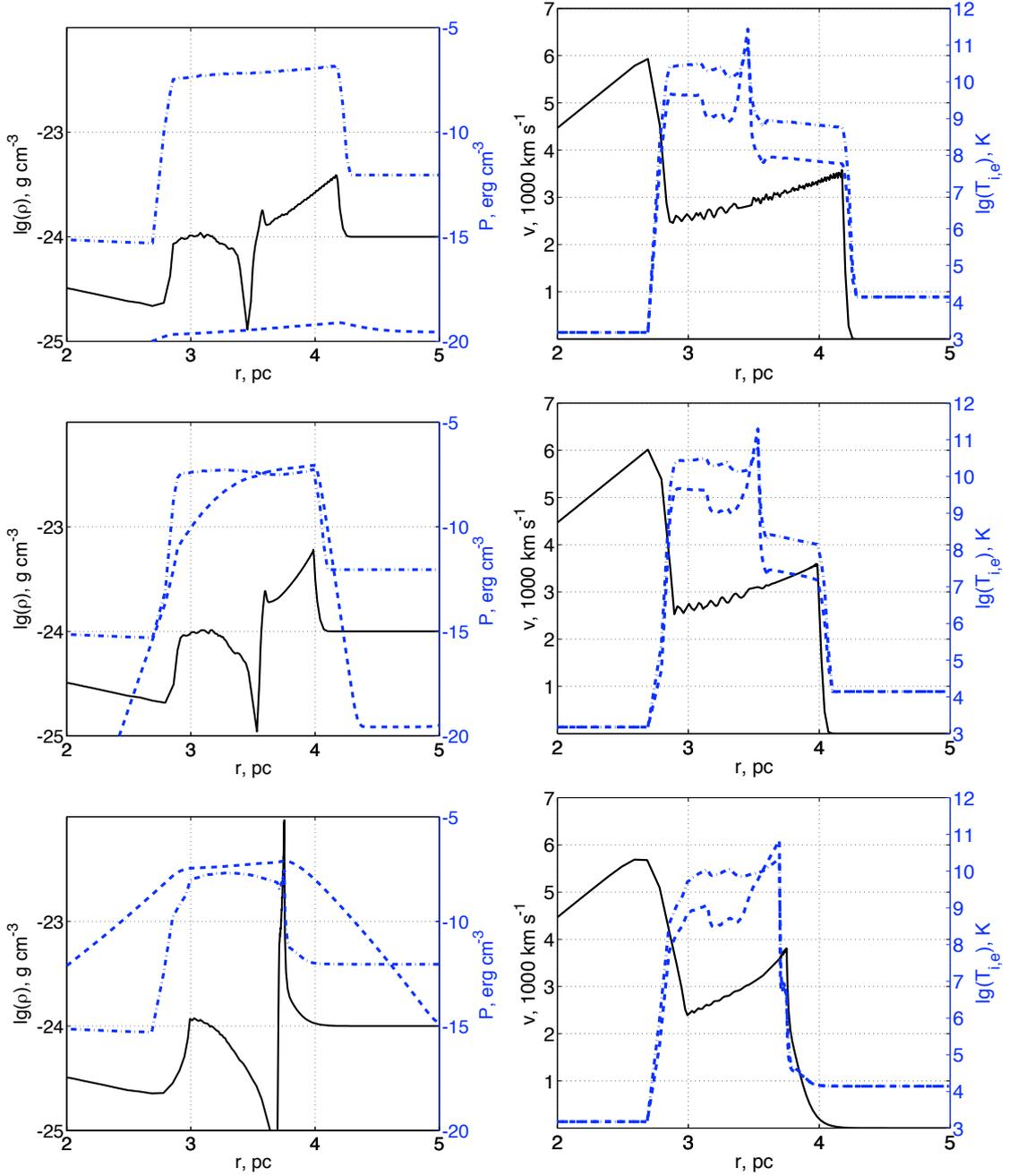
\FIGsi c440 c050403m_200_1d24_440_k3d26qcr00fs_pd c050403m_200_1d24_440_k3d26qcr00fs_vt  c050403m_200_1d24_440_k1d25qcr70fs_pd c050403m_200_1d24_440_k1d25qcr70fs_vt c050403m_200_1d24_440_k1d26qcr99fs_pd c050403m_200_1d24_440_k1d26qcr99fs_vt  [width=0.4\hsize,  angle=0] HD profiles for the SNR models at $t = 440$ years with $\rho_\mt{CSM} = 10^{-24}\,\mt{g\,cm}^{-3}$.
Top row shows the simulation with $\qcr = 0.0$, middle row --- $\qcr = 0.7$, $\kcr = 10^{25}\kcrd$;  bottom row --- $\qcr = 0.99$, $\kcr = 10^{26}\kcrd$. Left panels: density (black solid line, scale at the left-hand side), ion pressure (blue dash-dotted line, scale at the right-hand side), CR pressure (blue dashed line, scale at the right-hand side). Right panels:  velocity profile (black solid line, scale at the left-hand side), electron temperature profiles (blue dashed line, scale at the right-hand side), ion temperature profiles (dashed-dotted line, scale at the right-hand side).
%height=0.25\vsize,

Physical parameters of another set (SN1006 case) were chosen to match the remnant of SN1006 conditions: an ambient homogeneous circumstellar matter of density $\rho_0 =4\times10^{-26}\;\mt{g\,cm}^{-3}$ and the age of 1000 years. Hydrodynamical profiles of a few of these models are presented in \rfig{c1d3}. 
%{\bf Note, that the thermal and CR pressure balance in the models is in agreement with the studies of \citet{webb95}.}

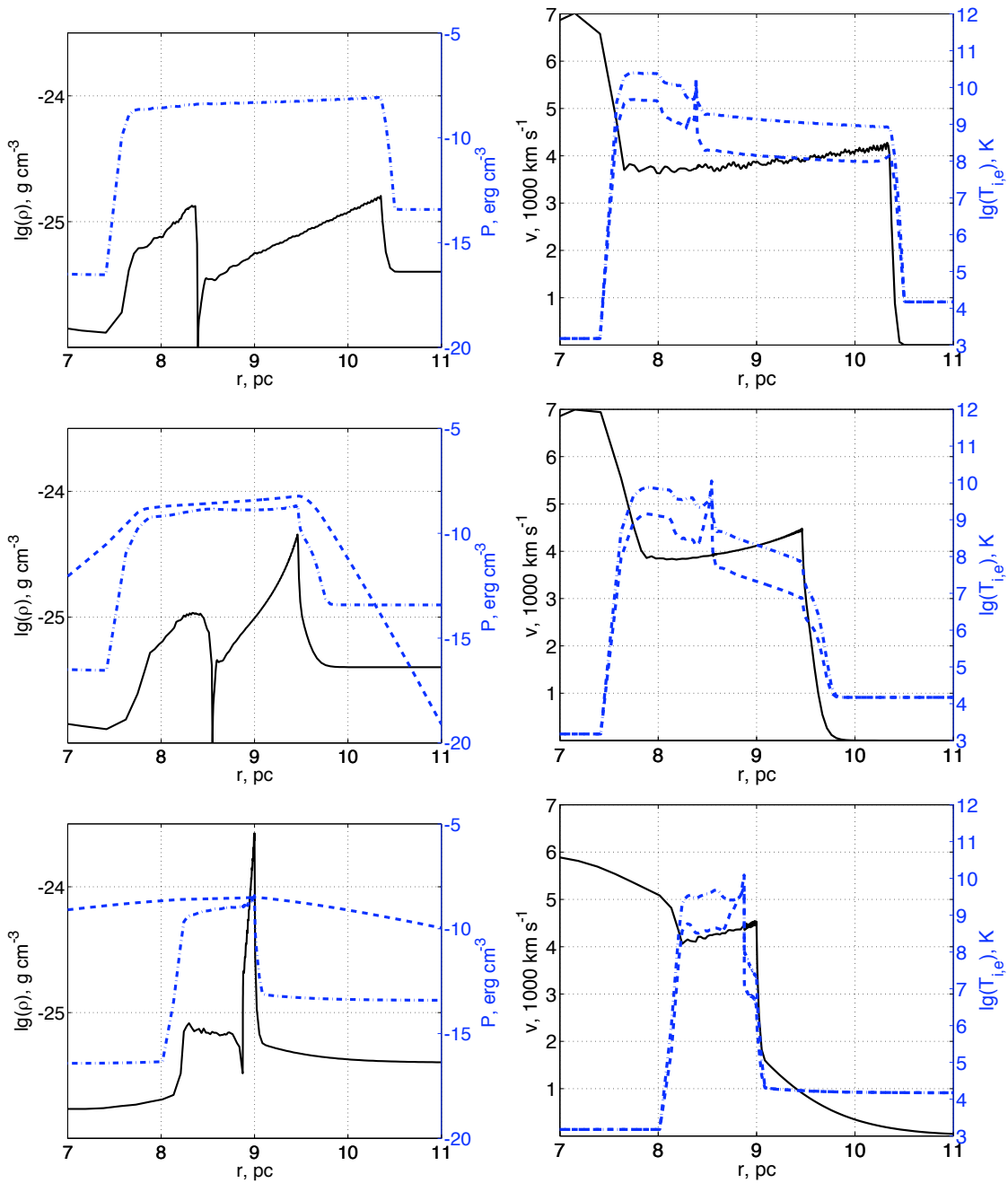
\FIGsi c1d3 c050403m_300_4d26_1d3_k1d24qcr00fs_pd c050403m_300_4d26_1d3_k1d24qcr00fs_vt  c050403m_300_4d26_1d3_k1d26qcr70fs_pd c050403m_300_4d26_1d3_k1d26qcr70fs_vt  c050403m_300_4d26_1d3_k1d27qcr99fs_pd c050403m_300_4d26_1d3_k1d27qcr99fs_vt   [width=0.4\hsize, angle=0] HD profiles for the SNR models at $t = 1000$ years with $\rho_\mt{CSM} = 4\times10^{-26}\,\mt{g\,cm}^{-3}$.
Top row shows the simulation with $\qcr = 0.0$, middle row --- $\qcr = 0.7$, $\kcr = 10^{25}\kcrd$;  bottom row --- $\qcr = 0.99$, $\kcr = 10^{26}\kcrd$. Left panels: density (black solid line, scale at the left-hand side), ion pressure (blue dash-dotted line, scale at the right-hand side), CR pressure (blue dashed line, scale at the right-hand side). Right panels:  velocity profile (black solid line, scale at the left-hand side), electron temperature profiles (blue dashed line, scale at the right-hand side), ion temperature profiles (dashed-dotted line, scale at the right-hand side).

The numerical models were verified to satisfy the Rankine-Hugoniot condition, derived for the system with energy losses \citep[e.g.][]{malkovdrury01, bykov08, vink10}. Namely the following relation was checked
\BE
\chi = {\gamma+1\over\gamma - \sqrt{1+2(\gamma^2-1)Q_\mt{esc}/(\rho_0v_\mt{S}^3)}}
\label{qesc}
\EE
where $\chi = \rho_\mt{max}/\rho_0$ is the compression ratio behind the shock, $\gamma$ --- adiabatic index, $Q_\mt{esc}$ --- energy lost by the system, $v_\mt{S}$  --- blast wave speed.  We rewrite the energy losses term from the energy flux conservation law \citep[e.g. formula (12) in][]{vink10} via a dimensionless parameter as
\BE
\ecr = 2Q_\mt{esc}/(\rho_0v_\mt{S}^3) = 1 - {1\over\chi^2} - {E + P\over\rho v_\mt{S}^2/2}
\label{eps}
\EE
where $E, P, \rho$ are total energy density, pressure and matter density behind the shock.

The top panel of \rfig{RHevol} shows the relation (\ref{qesc}) for $\gamma = 5/3$ --- solid line and $\gamma = 4/3$  --- dashed line. Data from the numerical models are presented with colored circles. Hue reflects the efficiency of CR acceleration, so that red point correspond to the models with $\qcr = 0.1$ and cyan --- to the models with $\qcr=0.99$.

The bottom panel of \rfig{RHevol} shows evolution of the compression ratio and ratio of the contact discontinuity (CD) to blast wave (BW) radii. The profiles are created from the Tycho's case models with $\rho_0 = 10^{-24}\,\mt{g\,cm}^{-3}$. Dashed lines show the simulation with $\qcr = 0.7$, solid lines --- $\qcr = 0.9$. Different colors correspond to different values of $\kcr$ (see legend). The curves are compared to those from \citet{volk08}. We see that $\chi$ profiles differ from the curves presented for Tycho's SNR in \citet[Fig. 1 in ][]{volk08}. In our approach the compression ratio is an increasing function of time, whereas in \citet{volk08} $\chi$ decreases with time. Radii ratio in our cases also does not follow exactly the same trend as in \citet{volk08}.

This divergence in the results can be explained by the magnetic field behind the shock. \citet{volk05} show that on average the downstream magnetic field pressure $B^2/(8\pi)$ is proportional to preshock gas ram pressure $\rho_0v_\mt{S}^2$. The ram pressure and the magnetic field dilutes as the remnant expands, thus the acceleration decreases. In our models, the acceleration efficiency depends only on the ram pressure (Eq.~\ref{artvisq}) and does not take into account the vanishing magnetic field. Thus with a constant $\qcr$ parameter, we somewhat underestimate the efficiency of the CR acceleration at the earlier times of the evolution and we overestimate it  at the later times.

Note, that for the simulation with efficient acceleration $\qcr = 0.9$ and with diffusion coefficient $\kcr=10^{27}\kcrd$ (solid magenta line) at the age of more than 1000 years a density spike starts to form. Structure of such a spike is illustrated in the bottom rows of \rfig{c440} and \rfig{c1d3}. Similar spikes are also present in the CR dominated shocks in simulations of \citet{wagner09}. Nevertheless, this type of structure is produced only at the extreme values of CR parameters and is unlikely to be realized in nature.

The CR precursor can be traced in the hydrodynamical profiles, presented in \rfig{c440} and \rfig{c1d3}, but the precursor region in front of the blast wave is thin ($\sim10^{17}\;\mt{cm}$) and cannot be accurately resolved.

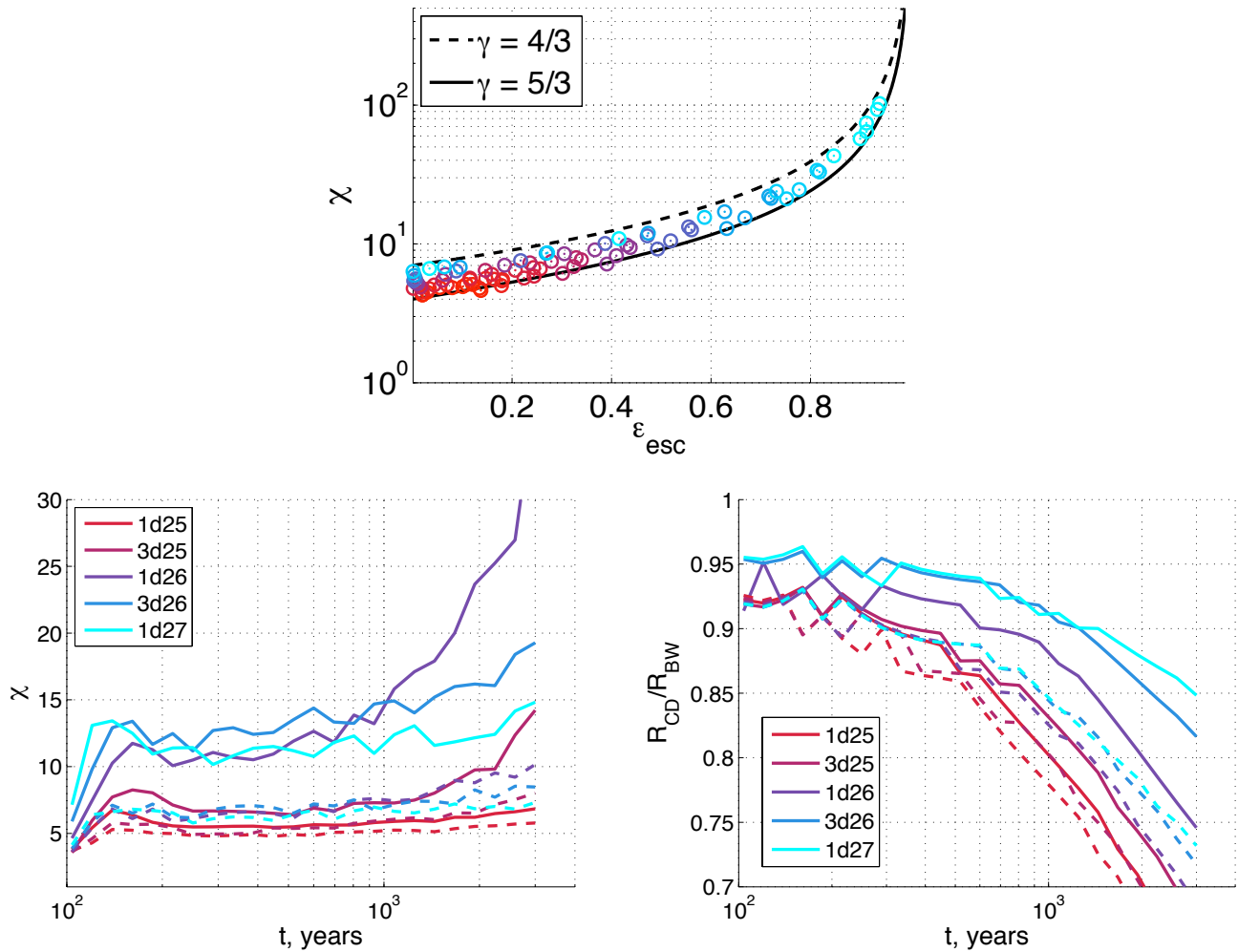
\FIGth RHevol RH_440_chi_esc chi_evol  r_evol [width=0.49\hsize,angle=0]  Top panel: Rankine-Hugoniot relation (\ref{qesc}) and data points from all the numerical models. Hue reflects the efficiency of CR acceleration. Bottom row: left panel shows evolution of compression ration $\chi$, right panel shows evolution of contact discontinuity to blast wave radii ratio, $\rho_0 = 10^{-24}\,\mt{g\,cm}^{-3}$. Dashed lines show models with $\qcr = 0.7$, solid lines --- $\qcr = 0.9$. Different colors indicate different values of the diffusion coefficient (see legend, where $\kcr$ values are given in units of $\kcrd$).

\section{Comparison with the Galactic SNRs}
\label{obs}
There is an evidence that the shocks of Tycho \citep[e.g.][]{warren05} and SN1006 \citep{chenai08, orlando08, miceli09} are modified by acceleration and diffusion of CR particles. Thus, we applied our numerical models to these Galactic supernova remnants. Locations of the blast wave (BW), the contact discontinuity (CD) and the reverse shock (RS) measured by \citet{warren05} for Tycho's SNR, were compared with the corresponding values obtained in our simulations.

The right plot of \rfig{rbw} shows the measured (horizontal strip) radii ratios and also results of the simulations (asterisks). To account for the three-dimentional projection effects and Rayleigh-Taylor (R-T) instabilities\footnote{Note, that the R-T instability is not considerably affected by acceleration at the forward shock \citep{ferrand10,wang10}}, that can produce fingers of ejecta that protrude out toward the BW \citep{chevalier92, wangchevalier01} we adopt a correction factor of $1.07$ \citep[][and references therein]{warren05, chenai08, volk08} to the modeled CD:BW radii ratios. The models that match the observed radii are boldfaced.

We also performed simulations with CR acceleration at the reverse shock with the same injection efficiency $\qcr$ and diffusion coefficient $\kcr$ as at the blast wave. In these models the layer of the shocked supernova ejecta becomes thinner, and the reverse shock approaches too close to the forward shock compared to the data from Tycho's SNR. Thus, we did not study these scenarios in details.

From an analysis of the data, presented in \rfig{rbw}, we derive that the most plausible models for Tycho are with $\qcr = 0.5-0.7$, $\kcr = 10^{24}-10^{25}\kcrd$.  In fact, models with $\qcr = 0.3$, $\kcr = 10^{26}\;\mt{cm}^2\mt{s}^{-1}$ and $\qcr = 0.9$, $\kcr = 10^{24}\kcrd$ also match the observation. Nevertheless, if we assume that diffusion coefficient $\kcr$ increases with the energy of the CR particles and that the amount of  energetic particles increases with the injection efficiency $\qcr$, then we conclude that low efficiency and high diffusion or high efficiency and low diffusion scenarios probably are not realized.

For Tycho, the simulations yield the compression ratio $\chi = 4.3-6.8$.
%(and $\chi = 4.6 - 8.5$ for R-T correction of 1.05).
These results are in agreement with \citet{chenai07} and \citet{volk08}.

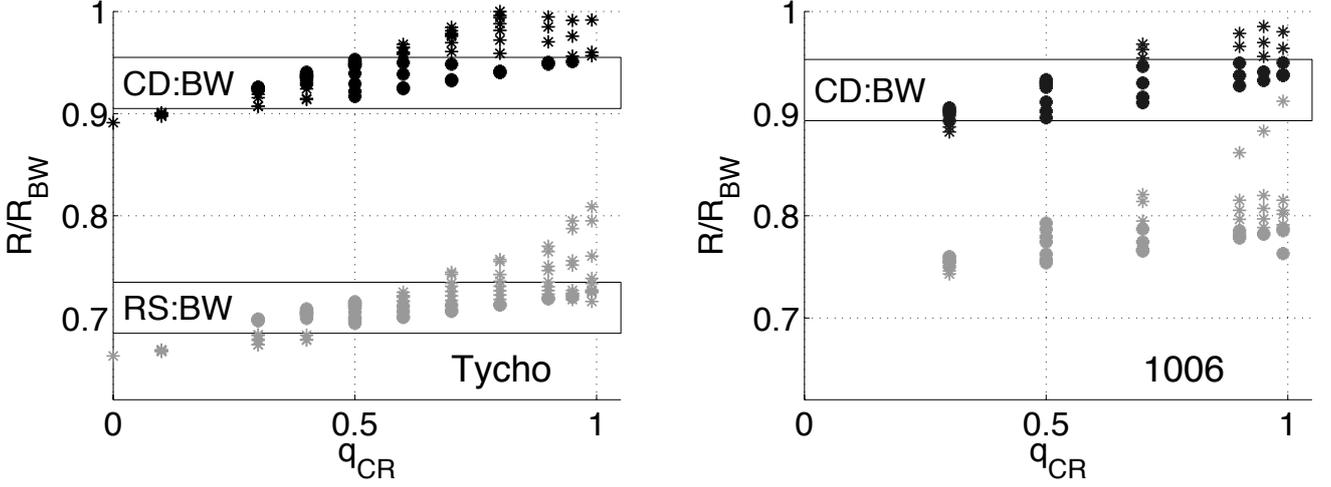
\FIGG rbw 440r107_chi43-68 1d3r107_chi47-83 [width=0.49\hsize,angle=0]  Measured and modeled radii ratios for Tycho's (left-hand panel) and SN1006 (right-hand panel) SNRs. $R_\mt{CD}/R_\mt{RW}$ (black symbols) and $R_\mt{RS}/R_\mt{RW}$ (grey symbols). Horizontal bars outline the measured by \citet{warren05}  and \citet{miceli09} values.

The similar comparison of the models to with SN1006 remnant data \citep{miceli09} is presented on the right plot of \rfig{rbw}. From these models we obtain the compression ratio behind the blast wave of the remnant $\chi = 4.7 - 8.3$. All the models with $\qcr=(0.3-0.9)$ match the observational data. Note, that models of SN1006 with constant adiabatic index $\gamma$ \citep{petruk10} could not explain the observed small distance between forward shock and contact discontinuity.

\section{Results}
\label{results}
We developed a hydro-code to simulate an evolution of a spherically-symmetrical supernova remnants with account for CR acceleration and time-dependent ionization. We created two sets of hydrodynamical models with different values of parameters for CR acceleration efficiency and CR energy losses. The comparison of these models with the measurement of BW:CD:RS radii of the Galactic SNRs suggests the following.

The simulations with $\qcr=(0.4-0.7)$ and $\kcr = (10^{24}-10^{25})\kcrd$ match the observed radii ratios of Tycho's SNR (\rfig{rbw}).  This range of values  covers the estimate $\kcr = 2\times10^{24}\;\kcrd$ found by \citet{wagner09}  for Tycho and are in agreement with the findings of \citet{parizot06} and \citet{eriksen11}. We derived the compression ratio of $\chi = (4.3-6.8)$, energy losses due to CR diffusion is of $\ecr = (0.1- 0.2)$ and distance to Tycho's SNR of $3.3$ pc \footnote{This value depends on the assumed explosion energy and CSM density (e.g. decrease of $\rho_0$ by a factor of three results in increase of the distance estimate by $25\%$).}. The distance is in agreement with the results of~\citet{hayato10} and \citet{tian10}.
%Errors in our distance estimates stem from the uncertainties of radial measurements and do not account for systematics.

For SN1006 remnant we obtained $\chi = (4.7-8.3)$ with losses of $\ecr = (0.2-0.5)$. We derived the distance to the SN1006 remnant as $2.0$ kpc.

\section{Discussion}
\label{discussion}

The observed ratios of SNR reverse shock, contact discontinuity and forward shock radii may depend on a number of yet unknown factors, such as structure of the CSM \citep[potential presupernova wind and density enhancement, see discussion in ][]{kosenko10,xu10}, CR acceleration and diffusion efficiency. Thus the results, presented in this study do not claim to be exhaustive, but rather describe one of the possible scenarios which explains the observations within the cosmic-ray acceleration paradigm.

In the numerical models  considered here, we do not turn on CR acceleration at the reverse shock. Results from the simulations with the same acceleration efficiency ($\qcr$) of the relativistic particles at the reverse shock do not match the observed $R_\mt{RS}/R_\mt{BW}$ ratio for Tycho's SNR. \citet{helder08, zirakashvili10} point out to the possible acceleration of particles by the reverse shock of CasA and SNR RX J1713.7--3946. In addition, numerical simulations of \citet{schure10} showed, that the reverse shock can re-accelerate particles, provided that the magnetic field is sufficiently strong. However, our models indicate that for  Tycho's SNR the cosmic rays acceleration at the reverse shock is not as efficient as at the forward shock.

The possible factors which suppress the acceleration at the reverse shock are as follows. The reverse shock velocity in the frame of ejecta is lower in comparison with the blast wave, and acceleration efficiency is probably an increasing function of the shock velocity.
%Also, the supernova ejecta consist of heavy particles (such as iron), which are more difficult to accelerate than protons of interstellar medium.
Moreover, the initial magnetic field of the progenitor white dwarf is diluted in the remnant by many orders of magnitude. Non-vanishing magnetic field is necessary for the acceleration to take place. Also, low efficiency of CR acceleration at the reverse shock was found for Kepler SNR by \citet{decourchelle00}.
%The possible explanation for this is twofold. The reverse shock velocity in the frame of ejecta is lower in comparison with the blast wave, and acceleration efficiency is an increasing function of the shock velocity. Moreover, the supernova ejecta consist of heavy particles (such as iron), which are more difficult to accelerate than protons of interstellar medium.

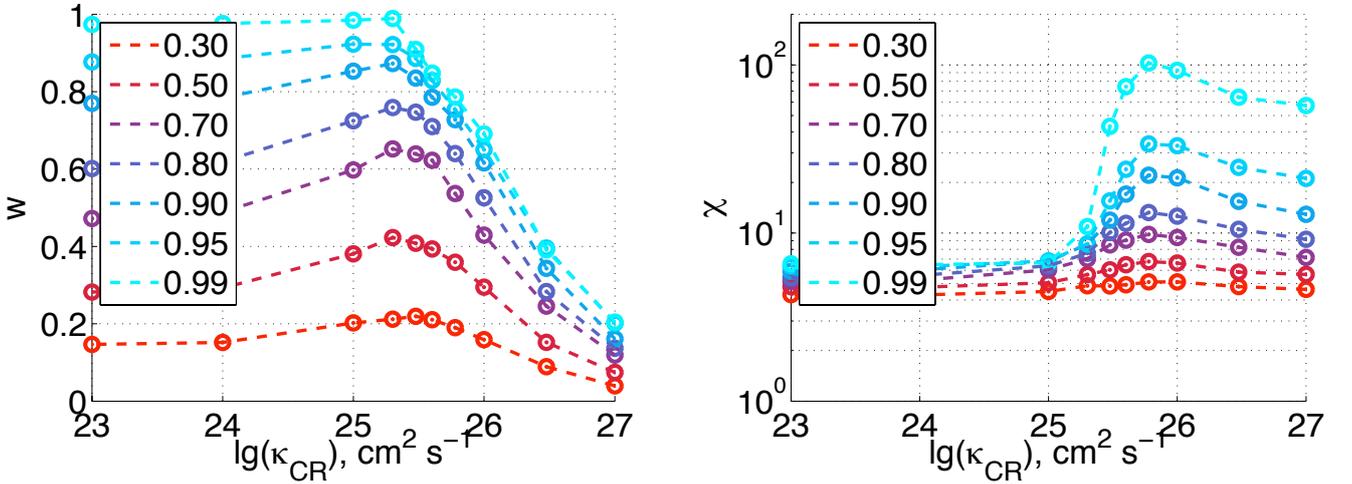
\FIGG Pcrq 440_w_kcr  440_chi_kcr  [width=0.49\hsize,angle=0]  Left panel: CR pressure versus diffusion coefficient. Right panel: compression ratio $\chi$ versus diffusion coefficient $\kcr$ for different values of $\qcr$ (Tycho's case).

We compared the input and output parameters of the simulation with Rankine-Hugoniot (RH) relations, derived in two-fluid approach for the steady-state situation and a plane-parralel geometry, presented in \citet{vink10}. On the one hand, the values of $\Pcr/P_\mt{tot}$ measured in the models directly behind the shock cannot be compared to $w$ defined in \citet{vink10}.
The diffusion coefficient $\kcr$ implements the CR energy losses and thus the CR pressure $\Pcr$ is a non-monotonic  function of $\kcr$ (\rfig{Pcrq}). Namely, the pressure decreases with $\kcr$ in the regime of efficient acceleration, where $\kcr>3\times10^{25}\kcrd$. This is opposite to the behavior of $w$ defined by \citet{vink10}. Possible explanation is that spherical symmetrical expansion and inward CR diffusion efficiently dilute the pressure, while in the plane-parallel case described by \citet{vink10} these effects are absent.

On the other hand, the parameter $\qcr$ sets the share of the CR pressure taken from the entropy pool created by the shock and in some sense {\it can} be attributed to $w$. \rfig{RHpcr} shows profiles of compression ration $\ecr$ (left panel) and CR energy losses $\chi$ (right panel) as a function of $w = \Pcr/P_\mt{tot}$ \citep[Fig. 1 in ][ for Mach number $M_0 = 500$]{vink10}. In the same panels we plot $\ecr$ and $\chi$  measured in the models versus the input parameter $\qcr$. The intensity of the data points corresponds to the value of $\kcr$: the lowest value corresponds to the black asterisks, the models with the highest value of $\kcr$ shaded in light gray.

The right-hand plot of \rfig{RHpcr} can be explained with the help of the right-hand plot of \rfig{Pcrq}, where compression ratios versus diffusion coefficient for different acceleration efficiencies $\qcr$ are presented. This plot shows that $\chi$ has a maximum at certain $\kcr$. These maximal values of $\chi$ for different $\qcr$ follow the RH curve, plotted with thick black line in the right plot of \rfig{RHpcr}. Respectively, the maximum attainable energy losses are skirted by the RH curves, presented in the left panel of \rfig{RHpcr}.

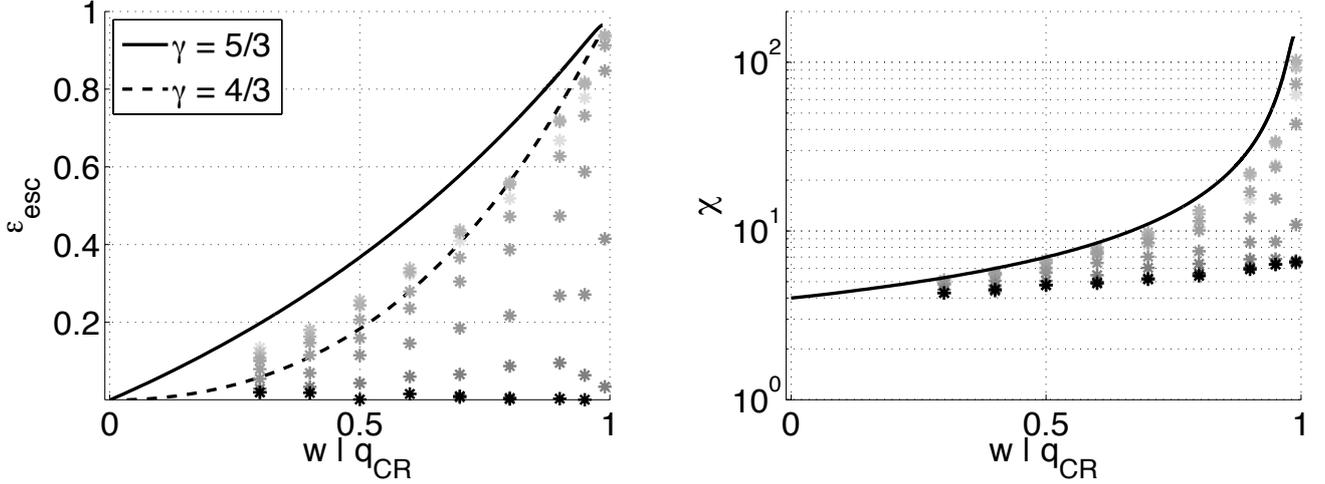
\FIGG RHpcr RH_440_esc_w  RH_440_chi_w  [width=0.49\hsize,angle=0]  Left panel:  escape CR energy versus $w$ and $\qcr$. Right panel:  compression ratio $\chi$ versus $w$ and $\qcr$ (Tycho's case). The Rankine-Hugoniot black curve corresponds for the case of Mach number $M_0 = 500$. Intensity of the data point is invers proportional to the values of diffusion coefficient $\kcr$, i.e. black asterisk correspond to $\kcr = 10^{24}\,\kcrd$, light grey correspond to $\kcr = 10^{27}\,\kcrd$.

The parameters $\qcr$ and $\kcr$ are fixed for each model in our method, while in reality they vary with time and location \citep[e.g.][]{zirakashvili10}. The effects of the variability of the diffusion coefficient was also discussed extensively by \citet{lagagecesarsky83}, where they point out that $\kcr$ increases away from the shock. Nevertheless, we assume that it is acceptable to use averaged constant values in this approach.

Even though we used two parameters for CR component in our models, in fact, the theory of diffusive CR acceleration assumes that, $\qcr$ and $\kcr$ are not independent. Thus it is possible to incorporate a physical relation between those two in future, reducing the problem to a case with only one free parameter.

At last we note, that in comparison of the models and the observations, we considered also a case with lower R-T correction factor of $1.05$. In this case the models that match the observed radii yield the compression ratio for Tycho's SNR systematically higher with $\chi = 4.6-8.5$, but within the uncertainties of our approach and errors of radii measurements, the final results are not altered considerably.

%Note, that at this stage we cannot compare directly our synthetic spectra to the observations, as there are too many parameters to match \citep[e.g.][]{badenes07}. The major uncertainty in spectral fitting is an supernova explosion model. Effects of the explosion mechanism on the resulting X-ray emission from a SNR are extensively discussed in \citet{badenes06,badenes08a,badenes08b}.
%\dk{Note that in \citet{zirakashvili10}  include self-consistently plasma kinetics and time-dependent ionization into their code.}

\section{Conclusion}
\label{conclusion}
We studied hydrodynamical models of supernova remnants evolution with account of diffusive cosmic-ray acceleration. We applied a two-fluid approach to investigate the effects of acceleration efficiency and energy losses on observable properties on the remnant. We compared the results of our simulations with the measured radii of BW, CD and RS of Tycho's and SN1006 SNRs.

We analyzed the numerical models and checked that the Rankine-Hugoniot relation for compression ratio and energy losses is met (\rfig{RHevol}). This method has a list of shortcomings but it includes all important relevant physical processes and easy to use for fast modeling of supernova remnants and for comparison with observations.

We found that, in order to explain radial properties of Tycho's SNR, the simulations yield the compression ratio behind the blast wave of $(4.3-6.8)$, energy losses due to cosmic ray particle diffusion of $(0.1- 0.2)\rho_0 v_\mt{S}^3/2$. Distance to Tycho is estimated as $3.3$ pc. For the case of SN1006 remnant, we found $\chi = (4.7-8.3)$ with losses of $(0.2-0.5)\,\rho_0 v_\mt{S}^3/2$. The distance to the remnant is $2.0$ kpc.

%In forthcoming studies we plan to investigate influence of the cosmic ray acceleration efficiency and losses on thermal X-ray emission calculated from the hydrodynamical models.
In the future we plan to employ the developed package for calculation of detailed thermal X-ray emission from the supernova remnant models modified by the acceleration. The resulting synthetic spectra will allows us to study the effects of different physical conditions of the shock plasma on the X-ray spectra. The analysis will be performed for different explosion models \citep[similar to the method of][]{badenes06, badenes08b} and later compared with observations of young remnants of type Ia supernova.

%There should not be a noticeable change in X-ray thermal spectra, as the temperature in the shocked ejecta is not affected by the CR parameters (we assume no acceleration by the reverse shock). Only temperature of the shocked CSM has dropped in the scenario with effective CR acceleration, thus we may expect, that this should result in different behavior of the bremhstralung /continuum emission.

\begin{acknowledgements}
DK is supported by an ``open competition'' grant and JV is supported
by a Vidi grant  from the Netherlands Organization for Scientific Research
(PI J.~Vink). The work of SB has been supported in part by World Premier International
Research Center Initiative, MEXT, Japan,
by Agency for Science and Innovations, Russia, contract 02.740.11.0250,
and by grants RFBR  10-02-00249-a, Sci.~Schools 3458.2010.2, 3899.2010.2,
and IZ73Z0-128180/1 of the Swiss National Science Foundation (SCOPES).
First of all, we thank the anonymous referee for valuable comments and discussions.
The authors are grateful to S.~Woosley for providing the models of thermonuclear supernovae.
We thank A.~Bykov and D.~Ellison for useful references and discussion and also  F.~Bocchino,  E.A.~Helder and K.~Schure.
\end{acknowledgements}

\begin{appendix}

\section{Basic equations}
\label{equations}

Originally  the method \citep[described by][]{sorokina04} solves the following system of hydrodynamical equations
%\dk{copy-paste from \citet{sorokina04}}
\begin{eqnarray}
 {\dr\over\dt}&=&u\,,\label{velo}\\
 {\dr\over\dm}&=&{1\over 4\pi r^2\rho}\,,\label{drm}\\
 {\du\over\dt}&=&\!
        {}-4\pi r^2\,{\d (P_e+P_i)\over\dm}-{Gm\over r^2}\,,
  \label{accel}\\
 \left({\dE_e\over\dT_e}\right)_\rho\,{\dT_e\over\dt}&=&\!
        {}-4\pi\,P_e\,{\d\over\dm}\left(r^2\,u\right)  \nonumber\\
        &&
          -\,4\pi\,{\d\over\dm}\left(r^2 F_{\rm cond}\right)
                                \nonumber\\
      &&-\eps_r\,
        -\,{\d\eps_{ion}\over\dt}
        -\left({\dE_e\over\dX_e}\right) {\dX_e\over\dt}
        \nonumber\\
        &&
        +\,{1\over\rho}\,\nu_{ie}\,
            k_b\left(T_i-T_e\right),
           \label{tpe}\\
 \left({\dE_i\over\dT_i}\right)_\rho\,{\dT_i\over\dt}&=&\!
        {}-4\pi\,P_i\,{\d\over\dm}\left(r^2\,u\right)  \nonumber\\
        &&
          -\,{1\over\rho}\,\nu_{ie}\,
            k_b\left(T_i-T_e\right),
                                \label{tpi}\\
 {\d{\mathbf X}\over\dt} &=&
        {}f(T_e,\rho,{\mathbf X}).
             \label{dotX}
\end{eqnarray}
Here $u$  is the velocity; $\rho$  is the density; $T_e$ and $T_i$
are the electron and ion temperatures; $P_e$ and $P_i$ are the respective
pressures, taking into account artificial viscosity (see description below); $E_e$ and $E_i$ are the thermal energies per unit mass of the gas element at the Lagrangean coordinate $m$ (corresponding to mass within a radius $r$) at the moment of time $t$;
$F_{\rm cond}$ is the energy flux due to the electron-electron and electron-ion thermal conduction;
$\nu_{ie}$ is the electron-ion collision frequency per unit volume;
 $\eps_r$ is the radiation energy loss rate per unit mass of the gas element;
$\d\eps_{ion}/\dt$ accounts for the change in specific thermal energy of gas
due to the change of ionization state; ${\mathbf X} = \{X_{\rm HI},X_{\rm HII},
X_{\rm HeI}, \dots,X_{\rm NiXXIX} \} $ is  the abundance vector  of all ions of all included elements relative to the total number of atoms and ions. Also $X_e=n_e/n_{b}$ for the number of electrons per baryon is introduced.
The physical processes described by these equations are discussed in \citet{sorokina04}.

\section{On the artificial viscosity}
\label{artviscapp}
Originally in {\sc supremna} the ion pressure was generated as $P_i=P_i(\mt{phys}) + q_i Q$, thus for the
electron pressure we put $P_e=P_e(\mt{phys}) + (1-q_i) Q$. Sign ``(phys)'' stands for physical %adiabatic
pressure in the system.

The meaning of this pressure can be clarified from the following simple thermodynamical relation.
The second law reads as $dE+PdV=0$, with $E$ --- internal energy, $P$ --- pressure, $dV$ --- volume element. Rewriting the pressure term as $P=P(\mt{phys}) + Q$, we get $dE+(P(\mt{phys})+Q)dV=0$. Thus, the term $Q$ plays a role of a source function in $dE+P(\mt{phys})dV=-QdV$. If there are energy losses $\eps$ in the system, then the final relation reads as $dE+PdV=-QdV+\eps$.

After the introduction of the CR component, using $\qcr$ parameter, we define $\Pcr = \Pcr (\mt{phys})+ \qcr Q$ ($\qcr = 0.0$ describes a case where the contribution from the cosmic-ray component is zero). Thus, we have the following distribution of the artificial viscosity (Eqn.~\ref{artvisq}) between ion and electron pressure:  $P_i = P_i({\rm phys})+(1 - \qcr) q_iQ$ and $P_e = P_e({\rm phys})+(1 - \qcr)(1 - q_i)Q$.  Therefore equation~(\ref{crtrans}) yields the source term in the form of $\Theta = \qcr Q (\du/\dr)$.
\end{appendix}

\bibliographystyle{aa}
\bibliography{crsupr}

\end{document}